\documentclass[a4paper]{jpconf}
\usepackage{bm,enumerate,dcolumn,tikz,graphicx,color,amsmath,amssymb}

\begin{document}
\title{ Atomic clocks and dark-matter signatures }

\author{Andrei Derevianko}

\address{Physics Department, University of Nevada, Reno, NV 89557, USA}

\ead{andrei@unr.edu}

\begin{abstract}
Recent developments in searches for dark-matter candidates with atomic clocks are reviewed. The intended audience is the atomic clock community.
\end{abstract}

\section{Dark matter problem and what we know about dark matter}
\label{Sec:Intro}
Multiple astrophysical observations suggest that the ordinary (luminous or baryonic) matter  (or simply ``us'') contributes only $\sim 5$\% to the total energy density budget of the Universe. Exacting the nature of the two other constituents, dark matter (DM) and dark energy,  is  a grand challenge to the contemporary  state of knowledge.
It is believed that dark matter is required for galaxy formations, while dark energy leads to the accelerated expansion of the Universe. The distinction between dark matter and dark energy can be formalized by treating them as cosmological fluids: they have different equations of state, dark matter is being pressure-less  while dark energy exerting negative pressure.  For further details I refer the reader to S. Weinberg  textbook on cosmology~\cite{WeinbergCosmologyBook} and also to reviews~\cite{Peebles2003,Bertone2005,Feng2010,Matarrese2011}. Below I  primarily focus on dark matter, although some aspects of the discussion, e.g., how an atomic clock may couple to various fields in a detectable way, would also apply to hypothetical dark energy fields.  

 All the evidence for dark matter (galactic rotation curves, gravitational lensing, peaks in the cosmic microwave background spectra, etc)  comes from galactic scale observations.  The challenge lies  in extrapolating down from the 10 kpc characteristic distances to the laboratory scales. This is a truly vast extrapolation scale and  a large number of theoretical  models can fit the observations.  The prevailing view is that all the theoretical constructs have to conform to the cold dark matter (CDM) model (for the purpose of the following discussion all DM objects move with velocities much smaller than the speed of light). More broadly, $\Lambda$-CDM model describes the large-scale structure formation  of the Universe~\cite{Blumenthal1984}.
 
 Our galaxy, the Milky Way, is embedded  into a DM halo and rotates through the halo. 
 Astrophysical simulations  provide  estimates of DM properties in the Solar system (see, e.g., \cite{NesSal13}). 
 The DM halo density distribution is usually parameterized by the Navarro-Frenk-White profile that is fit to various astrophysical observations.  Based on these  fits,  the DM energy density in the vicinity of Solar system is estimated to be
 $\rho_\mathrm{DM} \approx 0.3 \, \mathrm{GeV/cm^3}$, corresponding to about one hydrogen atoms per three cubic cm. Further, in the DM halo reference frame, the velocity distribution of DM objects is nearly Maxwellian with the dispersion of $v_\mathrm{vir} \sim 270\, \mathrm{km/s}$ (referred to as the virial velocity in the literature)
 and a sharp cut-off at the galactic escape velocity $v_\mathrm{esc} \approx 650  \, \mathrm{km/s}$.
Further, the Milky Way is a spiral galaxy  rotating through the DM halo. In particular, the Sun moves through the DM halo at galactic velocities $v_\mathrm{g} \approx 230\, \mathrm{km/s}$. 
 
  ``Dark''  universe does not absorb or emit electromagnetic radiation (dark particles are expected to carry no electric charge) and all the evidence is based on the gravitational interaction between the dark and luminous matter. It means that the gravitational interaction between DM objects and between DM and ordinary matter is assumed. But the gravity is weak on laboratory scales, so one needs to introduce additional interactions in order to make DM detectable by our instruments. Atomic clocks ultimately rely on locking electronic counters (time = (period of oscillation) $\times$ (number of oscillations))  to frequencies of atomic transitions. If DM affects atomic frequencies, then the DM effects can be potentially measured in terrestrial experiments. In Sec.~\ref{Sec:Portals}, I review a class of couplings to DM sector (``portals'') that translates into either transient or oscillating variations of fundamental constants. Such variations can pull on the transition frequencies and thereby affect atomic clocks.
 
  
 \begin{figure}[h]
\begin{center}
\begin{minipage}{17pc}

\includegraphics[width=14pc]{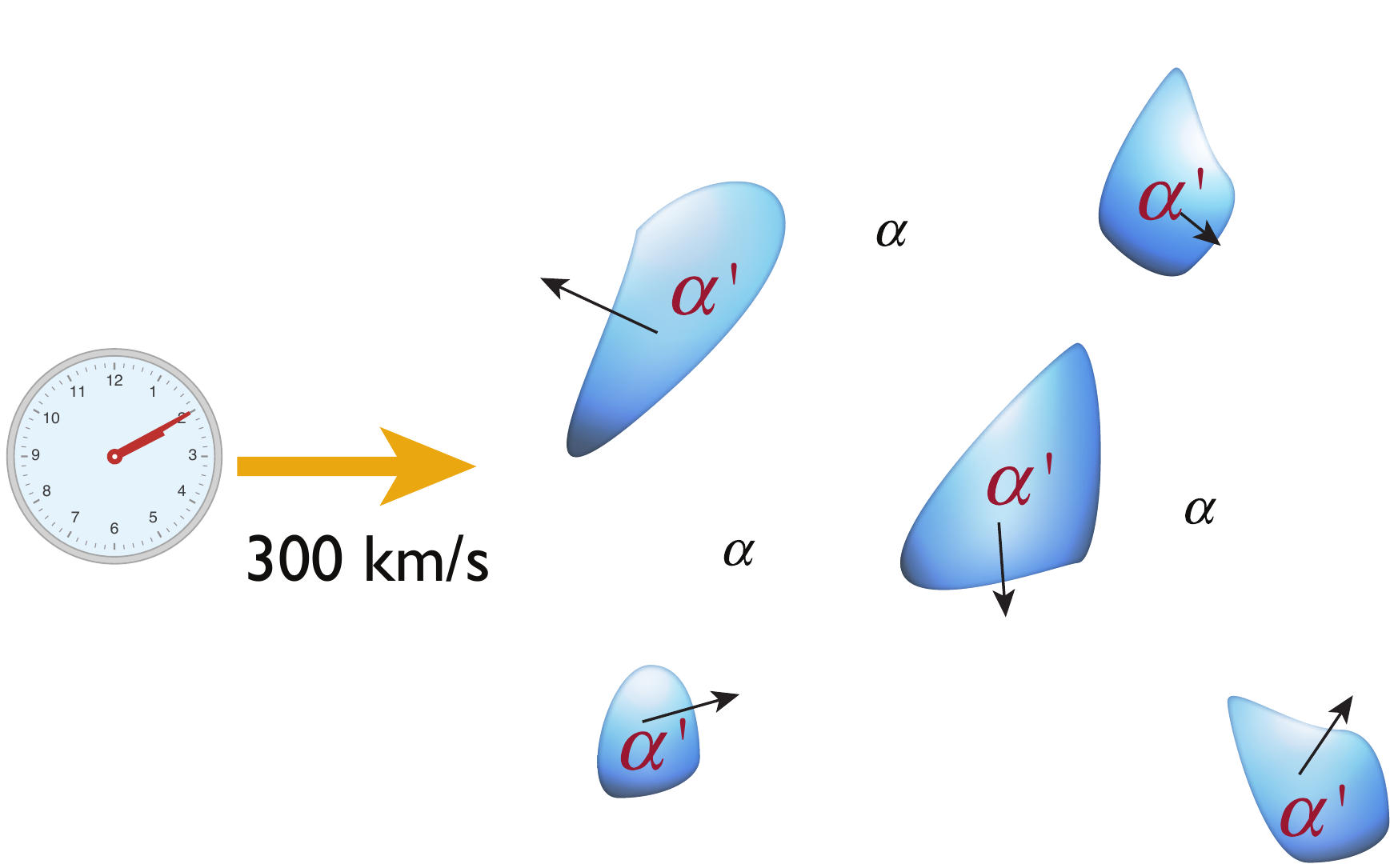}
\caption{\label{Fig:ClockMovingThroughDMhalo} (Colour online)
An atomic clock sweeps through the dark matter halo at galactic velocities.
Dark matter is assumed to be composed of extended objects (or clumps).
If there the difference of  fundamental constants (such as the fine-structure constant $\alpha$ in the figure) inside and outside the clumps, the clumps can cause the clock to slow down or speed up~\cite{DerPos14}.   }
\end{minipage}\hspace{2pc}%
\begin{minipage}{16pc}
\begin{center}
\includegraphics[width=10pc]{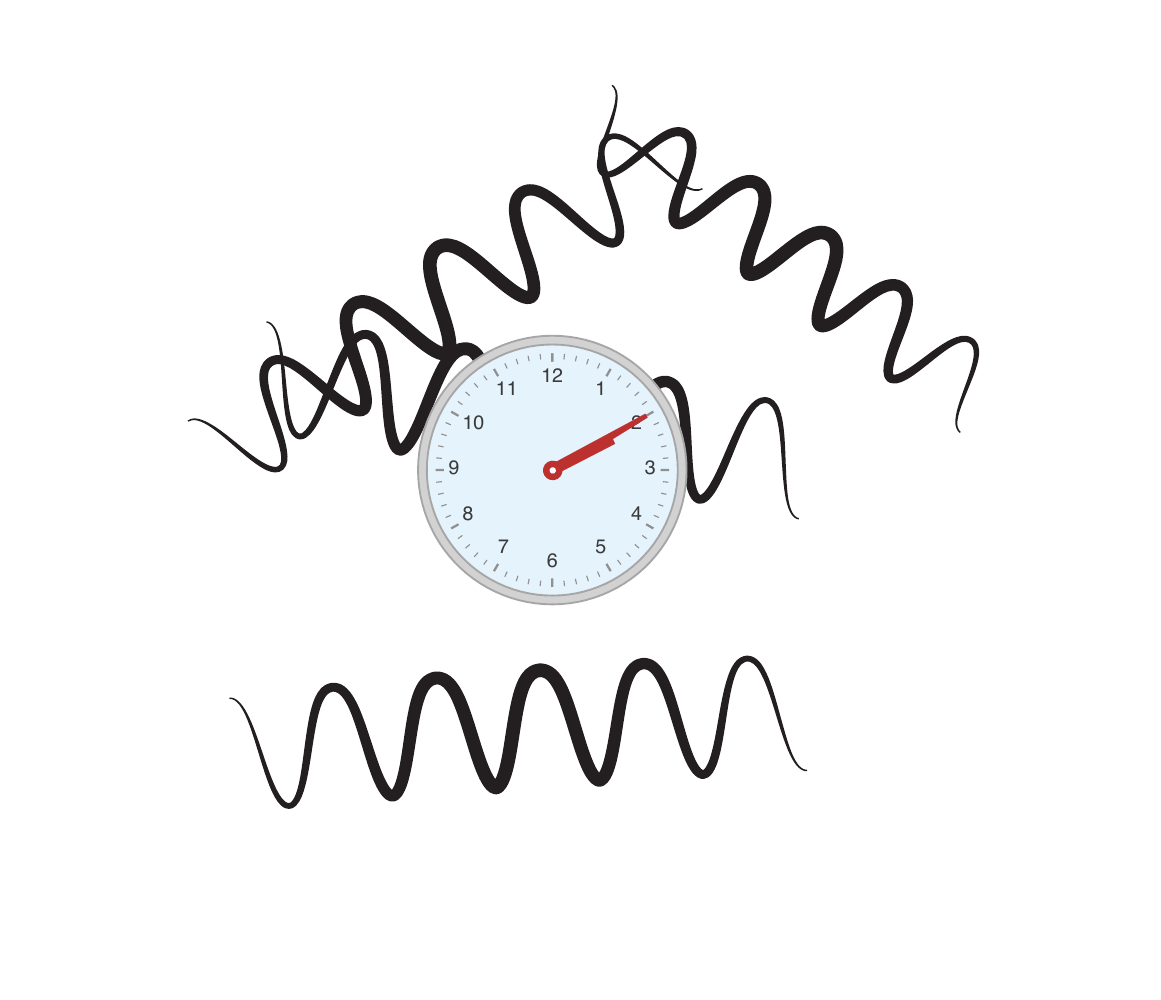}
\caption{\label{Fig:ClockInDilatonField} (Colour online) Ultra-light fields can lead to oscillating fundamental constants at the field Compton frequency. By Fourier-transforming a time series of clock frequency measurements, one could search for peaks in the power spectrum and potentially identify DM presence~\cite{ArvHuaTil15}.
}
\end{center}
\end{minipage} 
\end{center}
\end{figure}
 
Considering a wide variety of DM models, it is natural to ask if atomic clocks, being ultra-precise sensors, can be used to verify predictions of certain classes of DM models.  There were two recent proposals identifying such possibilities~\cite{DerPos14,ArvHuaTil15}. These two approaches lead to distinct DM signatures and are illustrated in Fig.~\ref{Fig:ClockMovingThroughDMhalo} and Fig.~\ref{Fig:ClockInDilatonField}.

Fig.~\ref{Fig:ClockMovingThroughDMhalo} depicts collisions of extended DM objects (or DM clumps) with an atomic clock traveling through the DM halo at galactic velocities. It is assumed that DM clumps effectively change the values of fundamental constants (such as the fine structure constant $\alpha$ and/or masses of elementary particles).   Such ideas based on correlated measurements using a network of atomic clocks were introduced in Ref.~\cite{DerPos14} and I review them in Sec.~\ref{Sec:TDM-signatures}. In the DM model of  Ref.~\cite{DerPos14}  the clumps are identified with topological defects (TD) formed due to the self-interaction of ultralight fields.  DM clumps can be formed in other models as well, so the TD model is not required per se. The question of {\em microstructure} of DM is an open question~\cite{Berezinsky2014}. From an observational point of view 
we can pose a  general question: can atomic clocks or other devices  detect our motion through the ``preferred'' DM halo reference frame?

 Another possibility (see Fig.~\ref{Fig:ClockInDilatonField}) is  DM  composed of ultralight fields, considered by Arvanitaki et al. \cite{ArvHuaTil15}. Such fields can lead to oscillating fundamental constants at the Compton frequency of the field. By Fourier-transforming a time series of clock frequency measurements, one could hunt for peaks in the power spectrum and potentially identify DM presence.

\section{Dark matter portals and new regimes of variation of fundamental constants}  
\label{Sec:Portals}
A systematic phenomenological approach to  DM-SM (Standard Model) sector couplings,  is  
that of the so-called  portals \cite{Essig:2013lka}, when the gauge-invariant operators of the SM
fields are coupled to the operators that contain DM fields. We focus on the SM-DM interactions in the form of the linear ($k =1$, or
dilaton) and quadratic ($k =2$) scalar portals,
\begin{equation}
-\mathcal{L}_{k}^\mathrm{int} =\phi ^{k} \left (\frac{m_{e} c^2 \bar{\psi }_{e} \psi _{e}}{%
\Lambda _{k ,e}^{k}} +\frac{m_{p} c^2 \bar{\psi }_{p} \psi _{p}}{\Lambda _{k
,p}^{k}} -\frac{1}{4\mu_0  \Lambda _{k ,\gamma }^{k}} F_{\mu \nu }^{2} +\ldots
\right )\; .  \label{Eq:Portals}
\end{equation}
The terms inside the brackets of Eq.~(\ref{Eq:Portals})  are  pieces from the SM sector Lagrangian density. These pieces are weighted with inverses of high-energy scales $\Lambda _{k ,X}$ which parametrize unknown coupling constants.
In particular, $m_{e ,p}$ and $\psi _{e ,p}$ are electron and proton masses and
fields ($\bar{\psi}=\psi^\dagger \gamma_0$), and $F_{\mu \nu }$ are the electromagnetic field tensor components. Linear portals were considered in Ref.~\cite{ArvHuaTil15} while quadratic portals --- in Ref.~\cite{DerPos14}. The energy scales are constrained from below from terrestrial experiments and astrophysical bounds. Generally such constraints  on the linear portals are far more stringent than those on the quadratic portals~\cite{Olive:2007aj}.

It is instructive to compare the portals~(\ref{Eq:Portals})  with the QED Lagrangian governing atomic physics:
\begin{equation}
-\mathcal{L}^\mathrm{QED} = -i \hbar c \bar{\psi}_e D \psi_e + m_e c^2 \bar{\psi}_e \psi_e + \frac{1}{4\mu_0} F^2_{\mu \nu} \, .
\end{equation}
It is clear that Eq.~(\ref{Eq:Portals})  when added to $\mathcal{L}^\mathrm{QED}$ leads to the 
modulation of fundamental constants by DM fields (notice that in the SI $\alpha = \mu_0 e^2 c/(4\pi \hbar$)):
\begin{equation}
m_{e ,p}^\mathrm{eff} =m_{e ,p} \times \left (1 +\frac{\phi (x ,t)^{k}}{\Lambda _{k ,e
,p}^{k}}\right ) , \text{\ \ }\alpha ^\mathrm{eff} \approx \alpha \times \left (1 +%
\frac{\phi (x ,t)^{k}}{\Lambda _{k ,\gamma }^{k}}\right )\; .
\label{Eq:TD:VarConst}
\end{equation}
Masses of quarks and other particles and various couplings (e.g., strong force) are affected in a similar way. 
\begin{figure}[h]
\includegraphics[width=18pc]{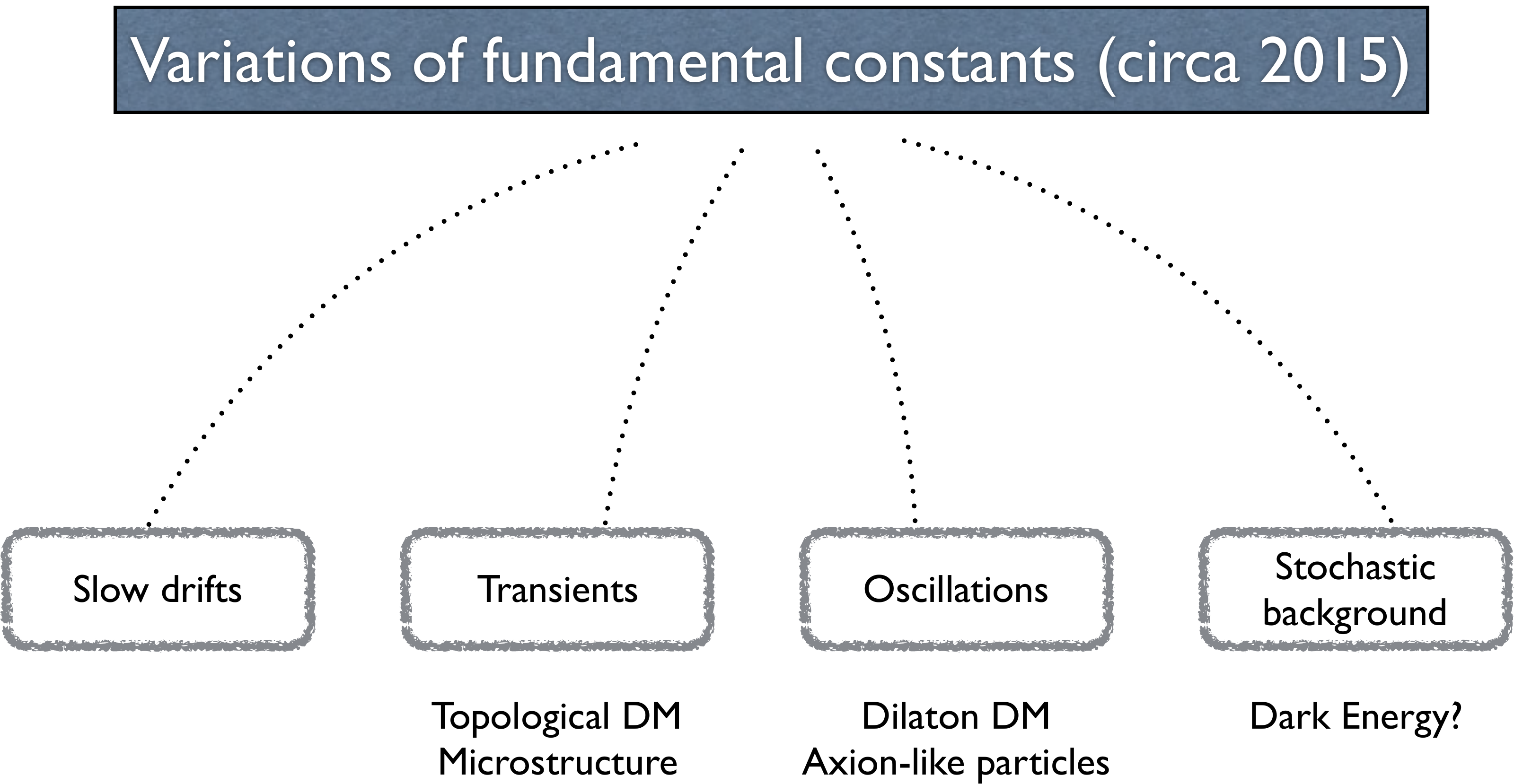}\hspace{2pc}%
\begin{minipage}[b]{14pc}\caption{\label{Fig:variations-new-regimes} (Colour online) Various scenarios of space-time variations of fundamental constants.}
\end{minipage}
\end{figure}

Now I invite the reader to take a closer look at the implications of Eq.~(\ref{Eq:TD:VarConst}).  For topological defects, outside the clump, by assumption, $\phi \rightarrow 0$ and
these portals renormalize masses and couplings only when the TD core overlaps with the quantum device. This,  via Eq.(\ref{Eq:TD:VarConst}),  leads to {\em transient} variation
of fundamental constants as the defect passes through the device.  The ultra-light fields oscillate at the Compton frequency, $\phi (
\boldsymbol{r} ,t) =A \cos (m_{\phi } c^{2} t -k_{\phi } \cdot
\boldsymbol{r} +\cdots )$, where $m_{\phi }$ is the mass associated with the
field and $k_{\phi}$ is the wave-vector of the field.  Thereby Eq.~(\ref{Eq:TD:VarConst}) leads to {\em oscillating} fundamental constants. 
 
The subject of space-time variation of fundamental constants has been explored previously  by the atomic clock community, however the attention until recently has focused on
the searches for slow drifts of fundamental constants~\cite{RosHumSch08}. 
As we see, DM searches with atomic clocks can be translated into searches for new regimes of space-time variation of fundamental constants. These various regimes are summarized in Fig.~\ref{Fig:variations-new-regimes}.

Operationally an atomic clock locks onto an atomic transition (frequency $f_0$). 
 One could parameterize the variation of the clock frequency in terms of  sensitivity coefficients $K_X$~\cite{FlaDzu09}
\begin{equation}
 \frac{\delta(f_0/U)}{ f_0/U} =\frac{\delta V}{V} \, , \, \,
 V =  \alpha^{K_\alpha} \left( \frac{ m_q}{\Lambda_\mathrm{QCD}} \right)^{K_q}  \left(\frac{m_e}{m_p}\right)^{K_{me/p}} \, . \label{Eq:DetailedCoeff}
\end{equation}
Here $U$ is the unit of  frequency, $m_q$ is a quark mass, $\Lambda_\mathrm{QCD}$ is the quantum chromodynamics mass scale, and $m_e/m_p$ is the electron to proton mass ratio.  In general, one distinguishes between two broad classes of atomic clocks: microwave and optical clocks.  
Microwave clocks, such as H, Rb and Cs, operate on hyperfine transitions; these depend both on $\Lambda_\mathrm{QCD}$ and $\alpha$. Nuclear-structure-dependent coefficient $K_q$ exhibits non-monotonic behavior~\cite{FlaTed06} and $K_{me/p}=1$ for hyperfine transitions. $K_\alpha$  grows with nuclear charge due to increasing  relativistic effects. 
  Optical clocks are only sensitive to the variation of $\alpha$.  It means that by comparing clocks of various sensitivity and type, one could discern individual terms in the portals~(\ref{Eq:Portals}).
  
\section{Relation to dark-matter models}
While the on-going particle physics DM searches focus on particles with masses $\sim 1-10^3 \, \mathrm{GeV}$,  here we consider an alternative: ultralight fields. Depending on the
initial field configuration at early cosmological times,  light fields
could lead to DM  oscillations about the minimum of their potential,
or form stable spatial configurations due
self-interaction potentials. The former
possibility leads to fields oscillating at Compton frequency (dilaton-type
DM~ \cite{ArvHuaTil15}) and the latter to the formation of topological
defects (TD) such as domain walls, strings and monopoles (TD-type DM~ \cite{DerPos14}). These two models were  introduced in Sec.~\ref{Sec:Intro}.

The maximum values of portal couplings~(\ref{Eq:Portals}) depend on the DM field amplitudes $A$;  these can related to the dark-matter
energy density in the Solar system neighborhood in the assumption that such models individually  saturate the DM energy density. For example, a gravitationally interacting gas of monopole TDs is a pressureless gas and can account for the observed properties of cold DM~\cite{DerPos14}.  The spatial extent of the monopole $d$ is given by the Compton wavelength $d\sim \hbar/(m_\phi c)$.  In this model, monopoles populate DM halo. As discussed in Sec.~\ref{Sec:Intro}, while the velocity distribution of monopoles is quasi-Maxwellian in the halo reference frame, 
the Earth is moving through the halo at galactic velocities $v_g \sim 300 \, \mathrm{km/s}$.  
The DM field amplitude inside the monopole can be expressed as
$A = \left( \frac{ \rho_\mathrm{DM} (\hbar c)^3}{m_\phi c^2} \, 
\frac{\mathcal{T}}{\hbar} \frac{v_g}{c} \right)^{1/2}$,  where $\mathcal{T}$ is the  average time between consecutive
collisions with the Earth (assuming $d \lesssim R_\oplus$, the Earth radius).  For dilaton-type DM,
$A=\frac{\hbar}{m_\phi c} \, \sqrt{ 2 \rho_\mathrm{DM} }$.

\section{Dark matter signatures} 
\label{Sec:TDM-signatures}
\begin{figure}[h]
\includegraphics[]{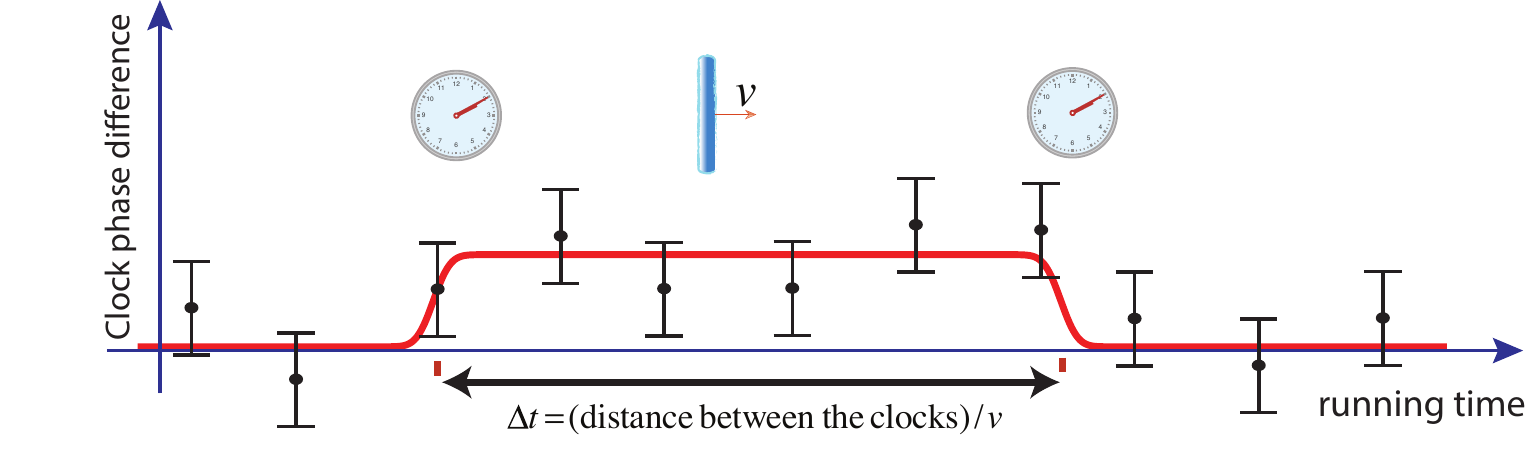}\hspace{2pc}%
\caption{\label{Fig:error-analysis-illistration} (Colour online)  Spatially-separated and initially-synchronized identical clocks are expected to  exhibit a distinct de-synchronization and re-synchronization pattern due to an encounter with a dark matter clump.}
\end{figure}

Inevitable clock noise (especially flicker noise) for a single clock can mimic an encounter with a DM clump.
Moreover, even if the DM induced clock glitches are large, an unsuspecting experimentalist is likely to discard the event and attribute it to something perhaps unexplained but mundane (see blog post~\cite{DerGroupBlog_whenWould}).
 In the  DM TD searches the solution~\cite{DerPos14} is to rely on a geographically distributed clock {\em network} which seeks synchronous propagation of clock ``glitches'' at galactic velocities through the network. This approach is similar to the  
proposed magnetometer GNOME network~\cite{Pustelny:2013rza} or gravitational wave detection~\cite{LIGOfirstObservation2016}.  In the context of clocks, two spatially-separated and initially-synchronized identical clocks are expected to exhibit a distinct de-synchronization and re-synchronization pattern, shown in Fig.~\ref{Fig:error-analysis-illistration}  for an encounter with a relatively thin (compared to the distance between the clocks) clump. The duration of the characteristic ``hump'' is given by $\Delta t= l/v$, with $l$ being the distance between the clocks and $v$ being the relative velocity of the encounter. 
If $v\sim v_g$, $\Delta t \sim 3 \, \mathrm{s}$ for a trans-continental network ($l \sim 1,000 \,\mathrm{km}$) and $\Delta t \sim  150 \, \mathrm{s}$ for clocks onboard navigational satellites ($l \sim 50,000 \, \mathrm{km}$).

As indicated in Fig.~\ref{Fig:error-analysis-illistration}, the measurements are carried out in the presence of clock noise characterized by the Allan variance $\sigma_y(T_m)$, where $T_m$ is the sampling interval. 
The signal-to-noise ratio~\cite{DerPos14} reads
\begin{equation}
   S/N =  \frac{ c \hbar  \rho_\mathrm{DM} \mathcal{T}  d^2 } 
    { T_m \sigma_y(T_m) \sqrt{  2 T_m v/ l }} 
     \sum_X K_X \Lambda_X^{-2} \, .
    \label{Eq:clockLimit}
\end{equation}
Here  $\rho_\mathrm{DM}$ is the DM energy density in the vicinity of the Solar system and $\mathcal{T}$ is the characteristic time between subsequent  DM encounters. The above S/N
ratio has been recently re-evaluated~\cite{CalIngLev2015}  taking into account time-transfer link noise. Notice that the derived S/N is for {\em two} identical clocks. The statistical confidence in the event can be improved dramatically with many nodes in the network and  if one could detect a synchronous  propagation of the sought signal through the network.

To fully disentangle the energy scales $\Lambda_\alpha$,  $\Lambda_q$,  and $\Lambda_{me/p}$ one would require at each node at least two microwave clocks and one optical clock or three microwave clocks of different types.  As we track transient variation of fundamental constants, it is sufficient to have identical clocks on different nodes. This differs from the search for a slow-drift-in-time of fundamental constants where a typical experiment (for example,~\cite{RosHumSch08}) uses two co-located clocks with different sensitivity coefficients.

The noise scales as $T_m^{3/2} \sigma_y(T_m)$.  Typically $\sigma_y(T_m)$ scales down as $1/\sqrt{T_m}$ with increasing $T_m$.  Then $S/N \propto 1/T_m$ and  it is beneficial to work with shorter measurement intervals.  The minimum time between consecutive measurements is determined by several factors: in lattice clocks~\cite{DerKat11}, this would be an atomic ensemble preparation  time  (about 1 second), in microwave  fountain clocks~\cite{WynWey05} it the time of flight across interrogation chamber (also about 1 second).   The lattice clocks \cite{DerKat11} may be  best suited for TDM search due to their stability and accuracy. The statistical advantage of lattice clocks comes from a large number of atoms being interrogated simultaneously. As the detection schemes would benefit from improved  short-term stability, we believe that the search would  greatly benefit from advances in  Heisenberg-limited time-keeping with entangled atoms.

What are the requirements for the network? Ideally, it should  be large and dense. Large in order to maximize the collision cross-section and rate with the defects and dense in order for the smallest DM clumps to be ``caught in the net''. For example, the distance between two nearby GPS satellites  in the same orbital plane is  $\sim 3 \times 10^4 \, \mathrm{km}$, meaning that this is the smallest size the GPS constellation network can be sensitive to.

If the DM  events are not observed, it could mean  that either the DM model is incorrect, or that the clocks are not sensitive enough to measure the effects. In the latter case, setting $S/N=1$ establishes the limits on the energy scales (or coupling constants) entering the portals~(\ref{Eq:Portals}). An example of such an analysis can be found in~\cite{DerPos14} for quadratic portals; it is clear that the existing constraints  on $\Lambda_X$ can be dramatically improved even with the existing GPS network of relatively low-accuracy atomic clocks.  While we (the GPS.DM observatory~\cite{GPS.DM_web})  have initiated efforts on mining a decade-long GPS data set for DM ``clump'' signatures,
the DM search sensitivity can be  improved further with a trans-continental network~\cite{Predehl2012,Lisdat2015} of high-accuracy laboratory clocks. As to the dilaton-type DM,
the first experimental bounds have already appeared~\cite{Tilburg2015}  and I hope that the nascent   dark matter searches with atomic clocks  discover invaluable DM signatures in terrestrial experiments. In any case they have a powerful potential of improving existing bounds on exotic interactions.


\ack 
This work was supported in part by the US National Science Foundation.

\section*{References}

\providecommand{\newblock}{}

\end{document}